\documentclass[aps,preprintnumbers,floatfix,article,amsmath,amssymb,floatfix,10pt,prd,superscriptaddress,nofootinbib,scrartcl]{revtex4-2}
\usepackage{bm}
\usepackage{amsfonts}
\usepackage{latexsym}
\usepackage[latin1]{inputenc}
\usepackage{graphicx}
\usepackage{amsmath}
\usepackage{palatino}
\usepackage{mathpazo}
\usepackage[british]{babel}
\usepackage{hhline}
\usepackage{multirow}
\usepackage{textcomp}
\usepackage{float}
\usepackage{booktabs}
\usepackage{dcolumn}
\usepackage{hhline}
\usepackage{multirow}
\usepackage{ragged2e}
\usepackage{hyperref}
\hypersetup{colorlinks,citecolor=blue}
\hypersetup{colorlinks=true,linkcolor=red,filecolor=magenta,    urlcolor=cyan}
\usepackage{amsmath}
\usepackage{xcolor}
\usepackage{epsfig}
\usepackage{caption}
\usepackage{subcaption}
\usepackage{commath}
\def\jnl@style{\it}
\def\aaref@jnl#1{{\jnl@style#1}}

\def\aaref@jnl#1{{\jnl@style#1}}

\def\aj{\aaref@jnl{AJ}}                   
\def\apj{\aaref@jnl{ApJ}}                 
\def\apjl{\aaref@jnl{ApJ}}                
\def\apjs{\aaref@jnl{ApJS}}               
\def\apss{\aaref@jnl{Ap\&SS}}             
\def\aap{\aaref@jnl{A\&A}}                
\def\aapr{\aaref@jnl{A\&A~Rev.}}          
\def\aaps{\aaref@jnl{A\&AS}}              
\def\mnras{\aaref@jnl{Mon.~Not.~Roy.~Astron.~Soc.}}             
\def\prd{\aaref@jnl{Phys.~Rev.~D}}        
\def\prc{\aaref@jnl{Phys.~Rev.~C}}  
\def\prl{\aaref@jnl{Phys.~Rev.~Lett.}}    
\def\qjras{\aaref@jnl{QJRAS}}             
\def\skytel{\aaref@jnl{S\&T}}             
\def\ssr{\aaref@jnl{Space~Sci.~Rev.}}     
\def\zap{\aaref@jnl{ZAp}}                 
\def\nat{\aaref@jnl{Nature}}              
\def\aplett{\aaref@jnl{Astrophys.~Lett.}} 
\def\apspr{\aaref@jnl{Astrophys.~Space~Phys.~Res.}} 
\def\physrep{\aaref@jnl{Phys.~Rep.}}      
\def\physscr{\aaref@jnl{Phys.~Scr}}       
\def\commat{\aaref@jnl{Comm.~Math.~Phys.}}              
\def\science{\aaref@jnl{Science}}               
\def\cqg{\aaref@jnl{Classical Quant.~Grav.}}            
\def\jpcs{\aaref@jnl{JPCS}}                                     
\def\ijmpd{\aaref@jnl{Int.~J.~Mod.~Phys.~D}}                    
\def\grg{\aaref@jnl{Gen.~Relat.~Gravit.}}               
\def\rpp{\aaref@jnl{Rep.~Prog.~Phys.}}          
\def\npa{\aaref@jnl{Nucl.~Phys.~A}}        
\def\lrr{\aaref@jnl{Living Rev.~Rel.}}                   
\def\jcap{\aaref@jnl{J.~Cosmology Astropart.~Phys.}}    
\def\rmp{\aaref@jnl{Rev.~Mod.~Phys.}}   
\def\epjc{\aaref@jnl{Eur.~Phys.~J.~C}} 
\def\plb{\aaref@jnl{~Phy.~Lett.~B}} 
\def\mpla{\aaref@jnl{Mod.~Phy.~Lett.~A}} 
\def\arxiv{\aaref@jnl{arxiv.org}}

\allowdisplaybreaks[1]

\begin{document}
\color{black}       
\title{\bf Dynamical Stability Analysis of Accelerating $f(T)$ Gravity Models}

\author{L.K. Duchaniya}
\email{duchaniya98@gmail.com}
\affiliation{Department of Mathematics,
Birla Institute of Technology and Science-Pilani, Hyderabad Campus,
Hyderabad-500078, India.}

\author{Santosh V Lohakare} 
\email{lohakaresv@gmail.com}
\affiliation{Department of Mathematics,
Birla Institute of Technology and Science-Pilani, Hyderabad Campus,
Hyderabad-500078, India.}

\author{B. Mishra}
\email{bivu@hyderabad.bits-pilani.ac.in}
\affiliation{Department of Mathematics,
Birla Institute of Technology and Science-Pilani, Hyderabad Campus,
Hyderabad-500078, India.}

\author{S.K. Tripathy}
\email{tripathy\_sunil@rediffmail.com}
\affiliation{Department of Physics, Indira Gandhi Institute of Technology, Sarang, Dhenkanal, Odisha-759146, India.}

\begin{abstract}
\textbf{Abstract}:  In this paper, we have emphasized the stability analysis of the accelerating cosmological models obtained in $f(T)$ gravity theory. The behaviour of the models based on the evolution of the equation of state parameter shows phantom-like behaviour at the present epoch. The scalar perturbation technique is used to create the perturbed evolution equations, and the stability of the models has been demonstrated. Also, we have performed the dynamical system analysis for both the models. In the two specific $f(T)$ gravity models, three critical points are obtained in each model. In each model, at least one critical point has been observed to be stable. 
 
\end{abstract}

\maketitle
\textbf{Keywords}:  Accelerating models, Dynamical stability, Scalar perturbation, Critical points. 

\section{Introduction} \label{SEC I}

The theoretical studies and cosmological observations of the Universe suggest that at early times Universe has passed through the inflationary stage and at late times in the accelerated phase. Theoretically, it can be achieved in two ways. In the first approach, the Universe content is to be altered with the introduction of additional fields, phantom scalar, canonical scalar, vector fields and so on \cite{Copeland06,Bassett06,Cai10}. The second approach is to modify the gravitational sector \cite{Capozziello11}. Usually, the modified gravitational theories are formulated by extending the Einstein-Hilbert action, which is curvature-based; however, another class of gravitational modification can be done by extending the action of equivalent torsional formulation of General Relativity (GR) e.g., the teleparallel equivalent of GR (TEGR)  \cite{Einstein28,Maluf94,Unzicker05,Aldrovandi13}. The framing of the modified theory of gravity that leads to second-order equations in the four-dimensional space-time can be started from TEGR. In GR, the Levi-Civita connection means curvature, but no torsion has been used whereas in teleparallelism, the Weitzenb$\ddot{o}$ck connection means torsion, but no curvature has been used \cite{Weitzenbock23}. In this framework, the dynamical objects are the four linearly independent tetrad fields that forms the orthogonal bases for the tangent space at each point of space-time. Also, the torsion tensor has been formed from the products of the first derivative of the tetrad. The $f(R)$ gravity is the simplest modification of GR and so also the $f(T)$ gravity, which is a simple modification of TEGR. It is well known that the action of general teleparallel gravity fails to satisfy local Lorentz invariance \cite{Sotiriou11}. In the formulation of $f(T)$ gravity \cite{Bengochea09, Ferraro08, Linder10}, one begins with pure tetrad teleparallel gravity and the spin connection is considered to vanish identically. So, the torsion tensor is effectively replaced by coefficients of the governing parameters, which are not tensors under local Lorentz transformations. The violation of local Lorentz symmetry has been ignored in TEGR as it does not affect the field equations, but this is an issue in $f(T)$ gravity. It is noteworthy to mention that the teleparallel gravity utilizes the teleparallel connection $\Gamma^{\rho}_{~~\mu \nu}$ \cite{Krssak19,Bahamonde21}, which has the torsion but vanishing curvature. Whereas in curvature based geometries, the Levi-Civita conncetion, $\tilde{\Gamma}^{\rho}_{~~\mu \nu}$ of the metric is used, which has non-vanishing curvature of space time. Both the connections are metric compatible.  \\ 

The problems related to expanding Universe and late time  acceleration are studied in $f(T)$ gravity \cite{Ferraro08,Bengochea09,Linder10,Myrzakulov11,Dent11}. In this gravity, Wu and Yu \cite{Wu10} have performed the dynamical system analysis in the power-law model. They have shown the stable de Sitter phase and unstable matter and radiation dominated phase. Hohmann et al. \cite{Hohmannl17} have shown that there are no trajectories, which would  start from an initial accelerating period, then decelerates and finally transition back to the  accelerating de Sitter phase. Bamba et al. \cite{Bamba12, Bamba12a} have shown the finite time future singularity. Another important study in $f(T)$ gravity is the Noether symmetry approach to find the exact solution to the given Lagrangian \cite{Wei12,Jamil12}. Zheng and Huang \cite{Zheng11} have compared their model framed through the power-law with the observational prediction of $\Lambda$CDM and dark energy models. At the same time, Setare and Mohammadipour \cite{Setare13} have  found the cosmological dynamical that obeys the cosmic history of $\Lambda$CDM. Based on TEGR, the phase space dynamical analysis has been performed for the teleparallel dark energy scenario \cite{Xu12}. Another aspect is the correspondence of $f(T)$ gravity with the holographic dark energy model in power-law  entropy correction \citep{Karami13}. The cosmographic parameters in the reconstructed $f(T)$ gravity shows the consistency with the $\Lambda$CDM model.  The logarithmic type $f(T)$ model does not allow to cross the phantom-divide, whereas in the combined logarithmic and exponential terms, the phantom-divide crossing can be realized \cite{Bamba11}. Wu and Yu \cite{Wu11} have also suggested two $f(T)$ models that realizes the phantom-divide crossings for the equation of state parameter. The constant torsion and specific conditions for the pressure, spherically symmetric solutions can be obtained in $f(T)$ gravity \cite{Daouda12}. The viability of $f(T)$ gravity models has been tested with the future measurement of Hubble expansion and the Monte Carlo analysis indicates the improvement of parameter space \cite{Basilakos18}. In Ref. \cite{Anagnostopoulos19}, the $f(T)$ gravity has been constrained with several data sources such as Pantheon supernovae sample, Hubble constant measurements cosmic microwave shift parameter, redshift-space distortion measurement. The tachyon inflation in teleparallel gravity discussed in \cite{Akbarieh19}. The consequence of the equivalence principle violation in the electromagnetic sector on $f(T)$ gravity have been discussed in Ref. \cite{Said20}. Hubble data has been used to reconstruct the $f(T)$ Lagrangian with the background cosmological parameters in Ref. \cite{Cai20,Briffa20} and the same approach was extended to growth rate data to constrain the values of the $f(T)$ Lagrangian \cite{Said21}. The free parameters of $f(T)$ gravity have been analysed from the combined observational data sets. The form of $f(T)$ has been reconstructed from the data driven equation of state parameter \cite{Ren21}.\\

Some of the issues related to $f(T)$ gravity theory have been raised by several authors over the years. Ong et al. \cite{Ong13} have pointed out the presence of superluminal propagating modes which can be reveled by the use of characteristics equation which governs the dynamics in $f(T)$ gravity. Izumi et al. \cite{Izumi14} have shown the inconsistency of Brans-Dicke type of teleparallel gravity extension and also opined that $f(T)$ gravity theory admits local acausality. In the covariant $f(T)$ gravity approach, Krssak and  Saridakis \cite{Krssak16} used an arbitrary tetrad in an arbitrary coordinate system along with the corresponding spin connection. This has been resulted in the same physically relevant field equations. Subsequently the fully invariant approach \cite{Krssak19} has removed the misconceptions on the local Lorentz invariance of teleparallel gravity and its generalizations. A detailed review on the teleprallel gravity from theory to cosmology can be seen in Ref. \cite{Bahamonde21}. In spite of several path breaking research, the physical nature of dark energy still remains to be unknown and the present cosmological observation has provided the information that $70\%$ of the Universe is filled with dark energy. Many dynamical dark energy models have been constructed to provide some specific cosmological scenarios and also, some modified gravitational theories are proposed in Ref. \cite{Sotiriou10,Cai16}. We wish to mention here that the dynamical system approach has been studied in the curvature based gravity \cite{Odintsov17,Oikonomou19} at length, but there is a need to further investigate in $f(T)$ gravity.  Here, we shall study the cosmological model in $f(T)$ gravity and the cosmological parameters would be constrained to verify the cosmic expansion and late-time acceleration. The paper is organised as: in Sec. \ref{SEC II}, the $f(T)$ gravity field equations and the dynamical and EoS parameters are shown. In Sec. \ref{SEC III}, we have presented two accelerating cosmological models with some physically viable form of $f(T)$ and analysed the dynamical parameters. In Sec. \ref{SEC IV}, the stability of the models are discussed in the scalar perturbation approach whereas in Sec. \ref{SEC V}, the dynamical system approach has been used. Finally, in Sec. \ref{SEC VI}, the conclusions of the models are given.

\section{Mathematical formalism of $f(T)$ gravity} \label{SEC II}

In the teleparallel framework, GR can be reformulated by using the tetrads as the dynamical variable in place of the metric tensor \cite{Einstein28}. The tetrad is a basis $\lbrace \textbf{e}_A({\textbf{x})\rbrace}$, where $A=0,1,2,3$, of vectors in the space-time. Each vector $e_A$ can be decomposed in a coordinate basis and produces the components $e_a^{\mu}$. Hence, the orthogonality condition becomes,
\begin{equation}\label{1}
g_{\mu \nu}=\eta_{AB} e_{\mu}^{A} e_{\nu}^{B},
\end{equation}
where $g_{\mu \nu}$ is the metric tensor and $\eta_{AB}=diag(1,-1,-1,-1)$. The co-frame $\lbrace \textbf{e}^A\rbrace$ would enable to obtain the invertible of Eq. \eqref{1} as, $e^{\mu}_Ae^B_{\mu}=\delta_A^B$. Eq. \eqref{1} enabled us to write, $e=det[e^A_{\mu}]=\sqrt{-g}$.  The connection in $f(T)$ gravity \cite{Weitzenbock23} can be defined as, $ \Gamma^{\lambda}_{\nu \mu}\equiv e^{\lambda}_{A} \partial_{\mu} e^{A}_{\nu}$ with the torsion tensor as,
\begin{equation}\label{2}
T^{\lambda}_{\mu \nu}\equiv\hat\Gamma^{\lambda}_{\nu \mu}-\hat\Gamma^{\lambda}_{\mu \nu}=e^{\lambda}_{A} \partial_{\mu} e^{A}_{\nu}-e^{\lambda}_{A} \partial_{\nu} e^{A}_{\mu}.
\end{equation}
The torsion scalar can be obtained from the contraction of the torsion tensor,
\begin{equation}\label{3}
T \equiv \frac{1}{4} T^{\rho \mu \nu} T_{\rho \mu \nu}+\frac{1}{2} T^{\rho \mu \nu} T_{\nu \mu \rho}-T_{\rho \mu}^{~~\rho} T^{\nu \mu}_{~~\nu}
\end{equation}
and the action of teleparallel gravity can be constructed from the teleparallel Lagrangian. The $f(T)$ gravity is to generalize $T$ to a function $T+f(T)$, whose action can be written as \cite{Anagnostopoulos19},  
\begin{equation}\label{4}
S = \frac{1}{16 \pi G}\int d^{4}xe[T+f(T)+\mathcal{L}_{m}],
\end{equation}

where $G$ be the gravitational constant and for completeness, we have included the total matter Lagrangian ($\mathcal{L}_{m}$). We shall consider the natural system, $\kappa ^2=8\pi G=c =1$. Varying the action \eqref{4} with respect to the vierbein, the gravitational field equations can be obtained as,
\begin{equation}\label{5}
e^{-1}\partial_{\mu}(e e^{\rho}_{A}S_{\rho}^{~\mu \nu})[1+f_{T}]+e^{\rho}_{A}S_{\rho}^{~\mu \nu}\partial_{\mu}(T)f_{TT}-e^{\lambda}_{A}T^{\rho}_{~\mu \lambda}S_{\rho}^{~ \nu\mu}[1+f_{T}]+\frac{1}{4}e^{\nu}_{A}[T+f(T)]=4 \pi G e^{\rho}_{A}T_{~\rho}^{~~\nu}
\end{equation}\\ 

Here on wards, we shall write $f=f(T)$ with $f_{T}$ and $f_{TT}$ be the first and second order derivative with respect to $T$. Also $ T_{~\rho}^{~~\nu} $ represents total matter energy-momentum tensor and the superpotential, $S_{\rho}^{~~\mu \nu}\equiv\frac{1}{2}(K^{\mu \nu}_{~~~\rho}+\delta^{\mu}_{\rho}T^{\alpha \nu}_{~~~\alpha}-\delta^{\nu}_{\rho}T^{\alpha \mu}_{~~~\alpha})$,  where the contortion tensor \cite{Hehl76},  $K^{\mu \nu}_{~~~\rho}\equiv \frac{1}{2}(T^{\nu \mu}_{~~~\rho}+T_{\rho}^{~~\mu \nu}-T^{\mu \nu}_{~~~\rho})$ . To study the cosmological scenario, we consider

\begin{equation}
ds^{2}=dt^{2}-a^{2}(t)\delta_{ij} dx^{i} dx^{j} \label{6}
\end{equation}
where $a(t)$ be the scale factor and  $ e^{A}_{\mu}\equiv diag(1,a(t),a(t),a(t))$. Now the field equations of $f(T)$ graity for FLRW space-time can be obtained as,  

\begin{eqnarray}
3H^2&=&8\pi G(\rho_m+\rho_r)-\frac{f}{2}+Tf_T \label{7}\\
\dot{H}&=&-\frac{4\pi G(\rho_{m}+\rho_{r}+p_{m}+p_{r})}{1+f_T+2Tf_{TT}} \label{8} 
\end{eqnarray}

Where the Hubble parameter, $H\equiv\frac{\dot{a}}{a}$ with overdot denotes the derivative with respect to cosmic time $t$. Moreover, $\rho_{m}$, $\rho_{r}$ represents the matter-energy density and radiation-energy density respectively whereas its corresponding pressure terms are $p_{m}$, $p_{r}$. The total energy-momentum tensor is comprising of the matter and radiation sector. Now the field equations of $f(T)$ gravity in the dark energy sector pressure and energy density can be obtained as,
\begin{eqnarray}
\rho_{de}&\equiv&\frac{1}{16 \pi G}\left[-f+2Tf_{T}\right] \label{9} \\
p_{de}&\equiv&-\frac{1}{16 \pi G}\left[\frac{-f+Tf_{T}-2T^{2}f_{TT}}{1+f_{T}+2Tf_{TT}}\right]\label{10} 
\end{eqnarray}
The torsion $T=-6H^2$ has been substituted in the above equations that represents the energy density $\rho_{de}$ and  pressure $p_{de}$. Further the effective equation of state (EoS) parameter for the dark energy sector can be obtained as,
\begin{equation}\label{11}
\omega=-1+\frac{\left(f_T+2Tf_{TT}\right)\left(-f+T+2Tf_T\right)}{(1+f_{T}+2Tf_{TT})(-f+2Tf_{T})}
\end{equation}

To explain the cosmic acceleration phenomena in a homogeneous and isotropic Universe, the matter component will have large negative pressure. Varieties of dark energy phenomenological models are available in the literature to find a suitable dark energy candidate. Though cosmological constant $\omega=-1$ has provided some relief however it also suffers from the fine-tuning problem. So, the time-dependent EoS parameter based models have been preferred for the dark energy component. Another preferred reason for this time dependent EoS parameter is its possible parametrization in redshift function and the prescribed values of $\omega$ from the cosmological observations.

\section{Cosmological Model} \label{SEC III}
In the action \eqref{4}, the $f(T)$ gravity takes the form $T+f(T)$ and the torsion $T$ can be expressed in Hubble parameter and so also the EoS parameter. In order to obtain the phantom-like behaviour of the Universe in $f(T)$ gravity, the equation of state parameter to be less than $-1$, the energy density increases over time and remains positive throughout. Also, the matter remains stable always in $f(T)$ gravity theory \cite{Behboodi12}. So to obtain the behaviour of EoS parameter, we choose a scale factor that comes from the cosmic solutions with the phenomenological quintom matter as,  $a(t)=\left(t^{2}+\frac{\alpha}{1-\nu}\right)^{\frac{1}{3(1-\nu)}}$, where  $\alpha$ and $\nu$ are  changeable values and $\nu\neq1$ \cite{Cai07,Karimzadeh19}. The Hubble parameter, $H(t)=\frac{\dot{a}}{a}=\frac{2t}{3t^{2}(1-\nu)+3\alpha}$. The free parameters $\nu$ and $\alpha$ can be adjusted to obtain the parameter in the desired range. Also, with the redshift function, $a(t)=\frac{1}{1+z}$, the Hubble parameter can be parametrized as,
\begin{equation}\label{12}
H(z)= \frac{2(1+z)^{3(1-\nu)}\sqrt{(1+z)^{3(\nu -1)}+\frac{\alpha}{\nu -1}}}{3(\nu -1)}
\end{equation}

\begin{figure}[H]
\centering
\includegraphics[width=85mm]{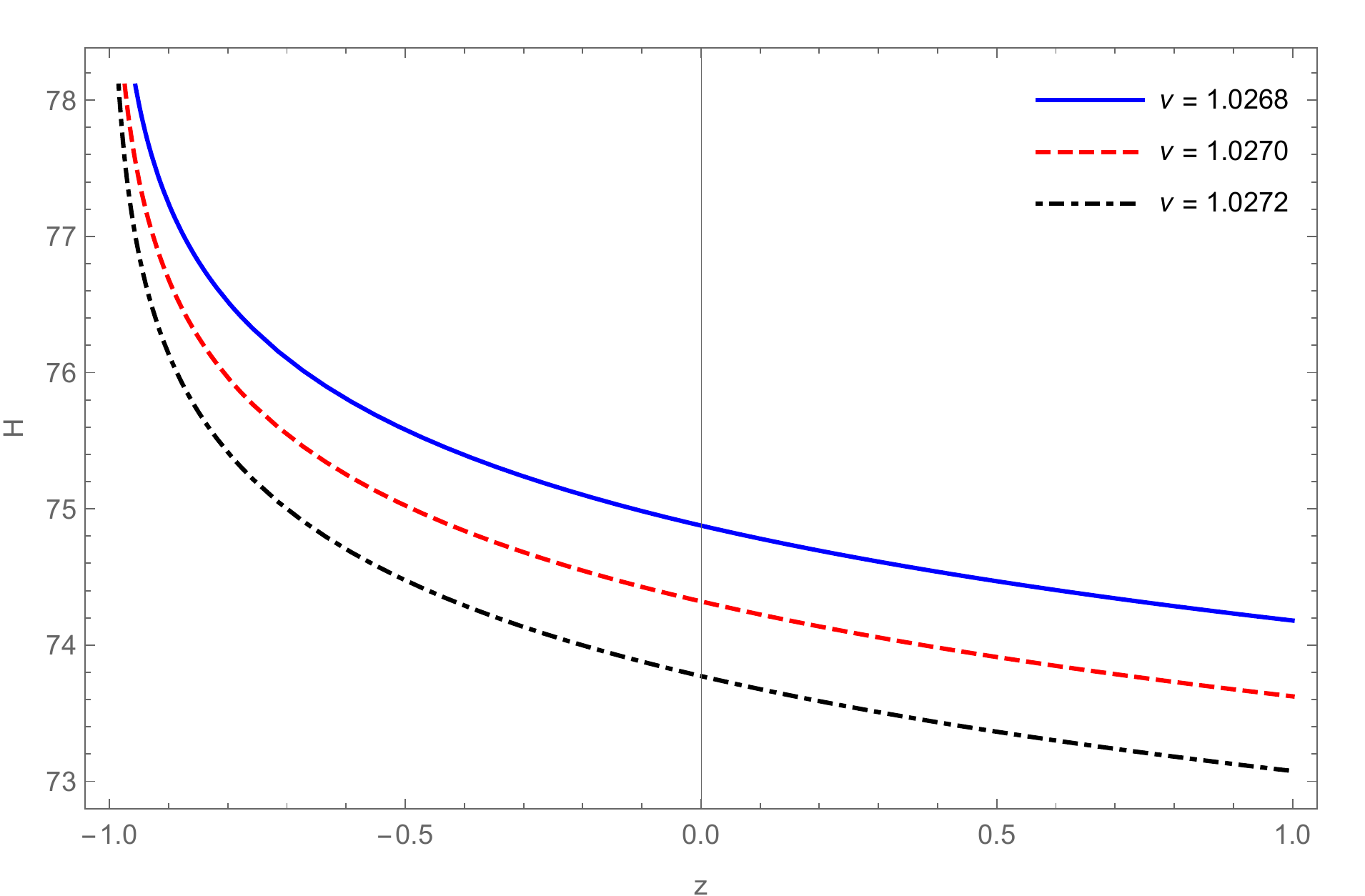}
\includegraphics[width=91mm]{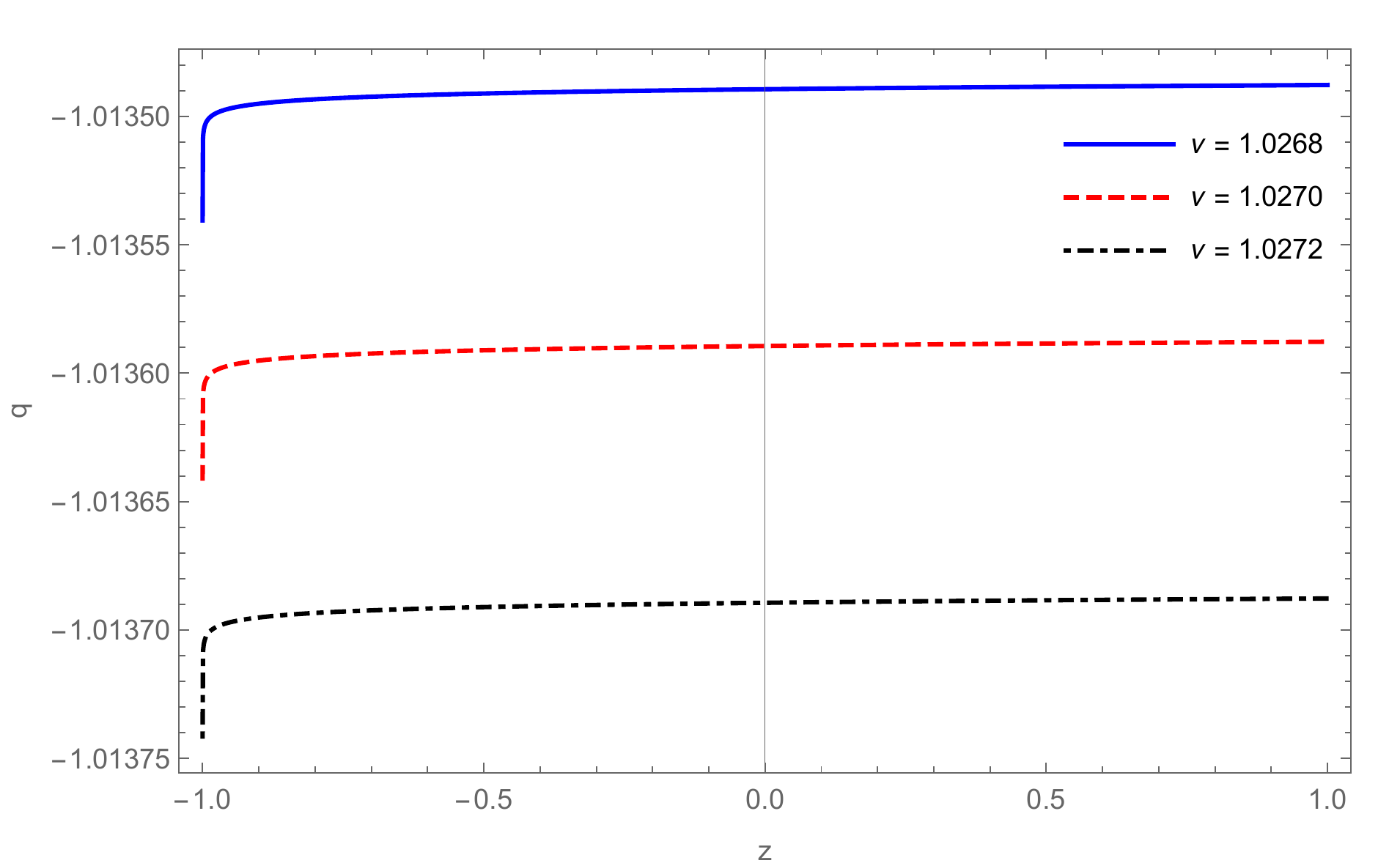}
\caption{Hubble parameter (left panel) and deceleration parameter (right panel) in redshift. The parameter scheme: $\alpha=0.21$.} \label{FIG1}
\end{figure}

The other important geometrical parameter is the deceleration parameter, $q = -1-\frac{3\alpha-3t^{2}(1-\nu)}{2t^{2}}$. The deceleration parameter decides the accelerating or decelerating behaviour of the Universe and the Hubble parameter provides the rate of expansion. The deceleration parameter can be parametrized as,
\begin{equation}\label{13}
q(z)=\frac{2 \alpha  (2-3\nu)  (z+1)^{3}+(\nu-1)(1-3 \nu) (z+1)^{3 \nu }}{2 \left(\alpha  (z+1)^{3}+(\nu -1) (z+1)^{3 \nu }\right)}.
\end{equation}
The graphical behaviour of the geometrical parameters are given in FIG. \ref{FIG1}. The Hubble parameter increases over time whereas the deceleration parameter decreases slowly. However, at late times it indicates a sudden dip. This behaviour of the deceleration parameter would show the accelerating era of the model. The present value of both the parameters are shown in Table-\ref{TABLE I} below and also compared with the corresponding values of the result of the cosmological observations. It has been observed that the value of the geometrical parameters obtained at present time are within the range of the cosmological observations.   

Now, we will discuss two examples pertaining to two form of $f(T)$. As a {\bf first example} of the cosmological evolution in $f(T)$ gravity, we will consider the case in which the function has the form \cite{Fortes21} as, $f(T)= T + \beta T^{2}$, where $\beta$ is constant. So that, $f_{T}=1+2\beta T$ and $f_{TT}=2 \beta $. Then Eqs. \eqref{9}-\eqref{11} reduce to,
\begin{eqnarray}
\rho_{de}&=&\left [\frac{-4  t^{2}(t^{2} (1-\nu) +\alpha)^{2} + 32 \beta t^{4}}{3(t^{2}(1-\nu)+\alpha)^{4}}\right]\label{14}\\
p_{de}&=&\left[\frac{16 \beta t^{4}}{3(t^{2} (1-\nu)+\alpha)^{4}-24 \beta t^{2}(t^{2}(1-\nu)+\alpha)^{2}}\right] \label{15}\\
\omega &=&\frac{4 \beta t^{2} (t^{2}(1-\nu)+\alpha)^{2}}{16 \beta t^{2}(t^{2}(1-\nu)+\alpha)^{2}-64 \beta^{2} t^{4}-(t^{2}(1-\nu)+\alpha)^{4}}\label{16}
\end{eqnarray}

\begin{figure}[H]
\centering
\includegraphics[width=85mm]{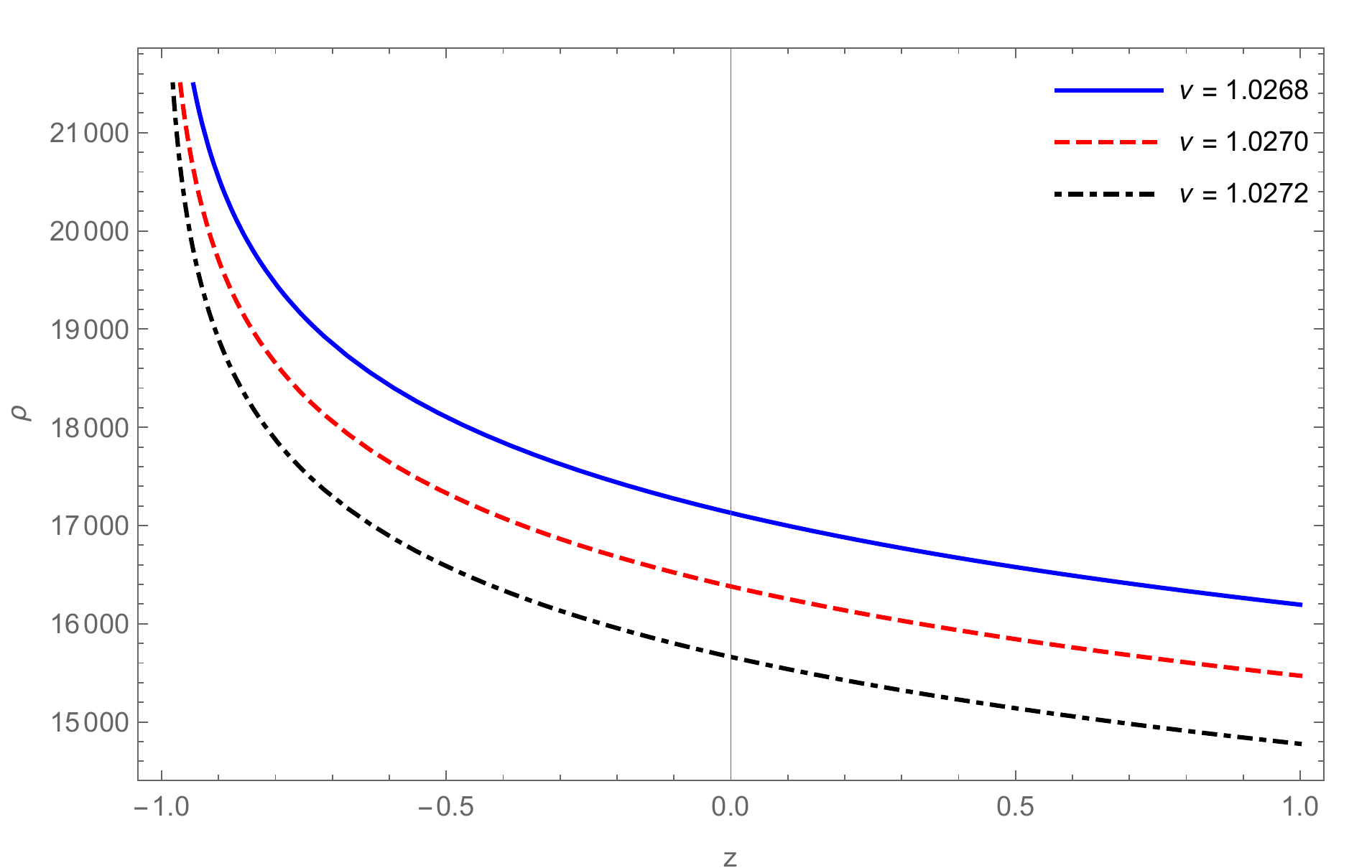}
\includegraphics[width=85mm]{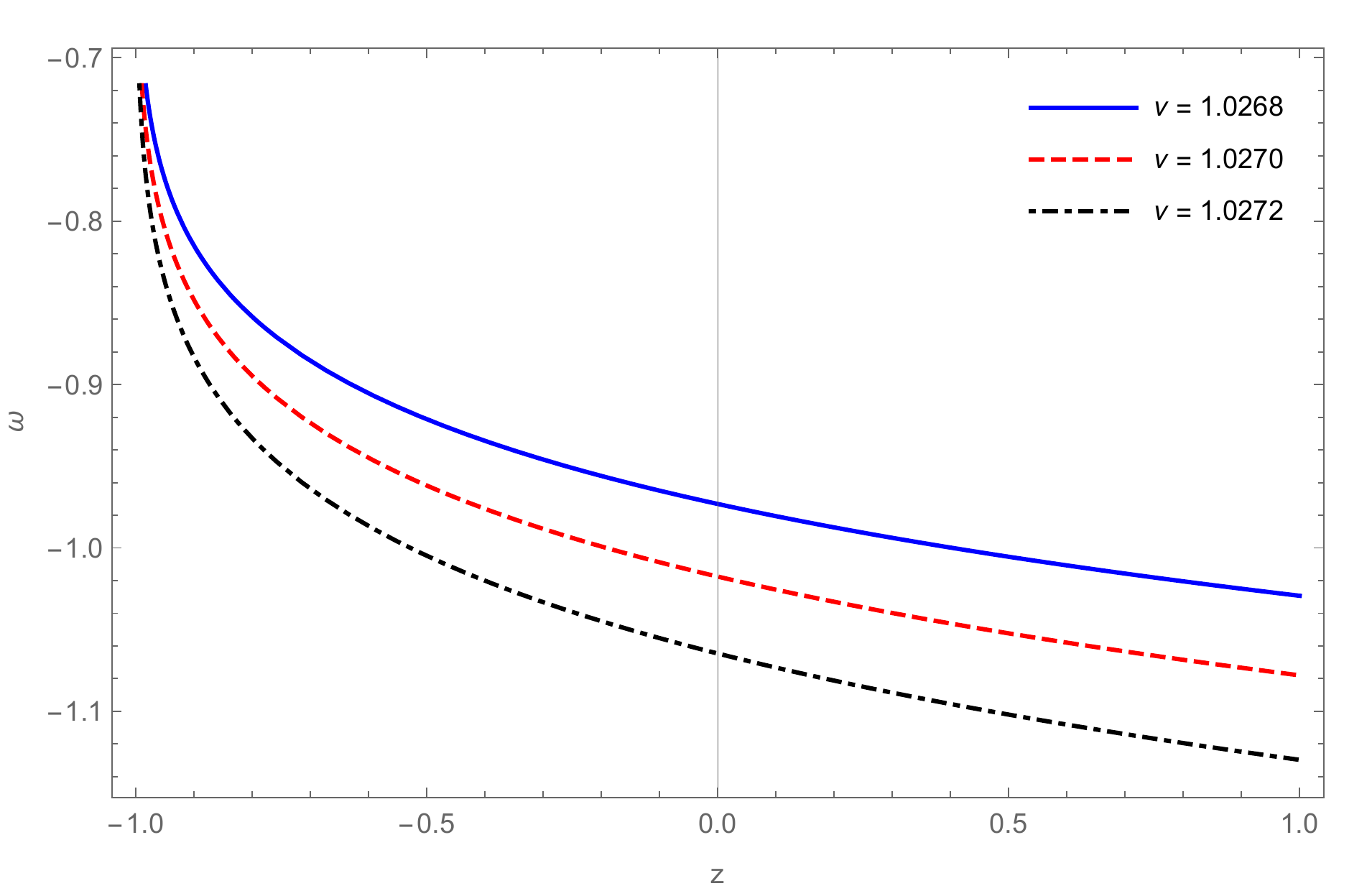}
\caption{Effective energy density (left panel) and EoS parameter (right panel) in redshift. The parameter scheme: $\alpha=0.21, \beta=0.0000208$.}
\label{FIG2}
\end{figure}

The evolutionary behaviour of the dynamical and EoS parameter is governed by the parameters $\alpha$, $\nu$ and $\beta$. First, the scale factor parameters are constrained to get the geometrical parameters in the desired range and now, in addition the model parameter has been constrained to obtain the positive energy density FIG. \ref{FIG2} (left panel). The energy density of dark energy increases over time and in late-times it has increased substantially. The EoS parameter has started the evolution from the phantom phase and at the late phase, it remains in the quintessence region. At the present time also it remains in the phantom phase and the present values are noted under Model I in  Table-\ref{TABLE I} for different $\nu$.\\  

As an {\bf second example}, we consider the form of $f(T)$ as, $f(T)=\left(\frac{T}{\lambda}\right)^{\gamma}$, where $\gamma$ and $\lambda$ are constants \cite{Rezazadeh16}. For this $f(T)$, we obtained $f_{T}=\frac{\gamma T^{\gamma-1}}{\lambda^{\gamma}}$ and $f_{TT}=\frac{\gamma (\gamma-1)T^{\gamma-2}}{\lambda^{\gamma}}$. Now, Eqs. \eqref{9}-\eqref{11} reduce to,  
\begin{eqnarray}
\rho_{de}&=&\frac{(2\gamma-1)}{2 \lambda^{\gamma}}\left[\frac{-24 t^{2}}{9 (t^{2}(1-\nu)+\alpha)^{2}}\right]^{\gamma} \label{17}\\
p_{de}&=&\frac{1}{2}\left[\frac{(2\gamma-1)(\gamma-1)(-24 t^{2})^{\gamma}}{\left(9 \lambda \right)^{\gamma}(t^{2}(1-\nu)+\alpha)^{2\gamma}+9(-24t^{2})^{\gamma-1}(t^{2}(1-\nu)+\alpha)^{2}\gamma (2\gamma-1)}\right]\label{18}\\
\omega&=&\frac{\left(9 \lambda \right)^{\gamma}(\gamma-1)(t^{2}(1-\nu)+\alpha)^{2\gamma}}{\left(9\lambda \right)^{\gamma}(t^{2}(1-\nu)+\alpha)^{2\gamma}+9(-24t^{2})^{\gamma-1}(t^{2}(1-\nu)+\alpha)^{2}\gamma (2\gamma-1)} \label{19}
\end{eqnarray}

\begin{figure}[H]
\centering
\includegraphics[width=85mm]{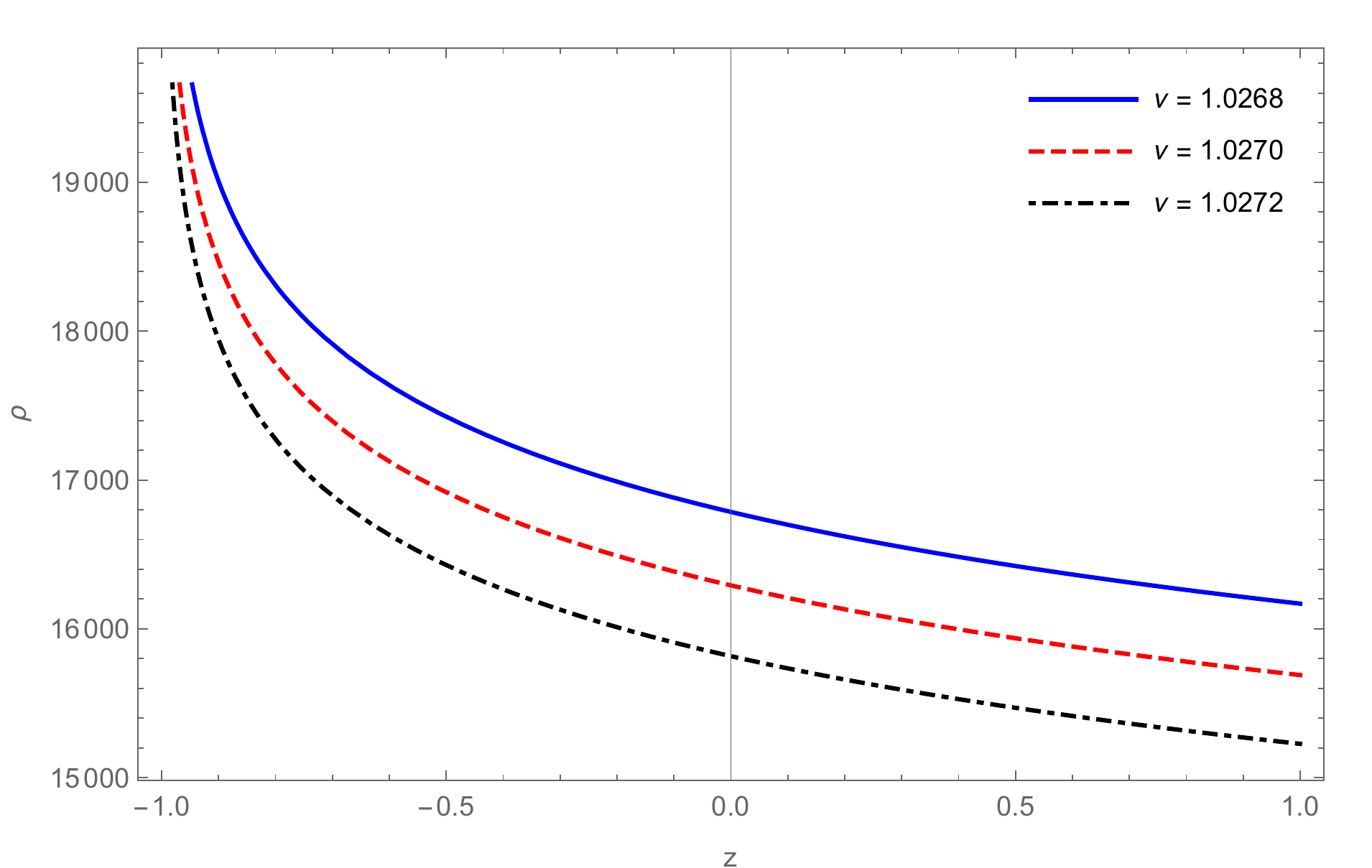}
\includegraphics[width=85mm]{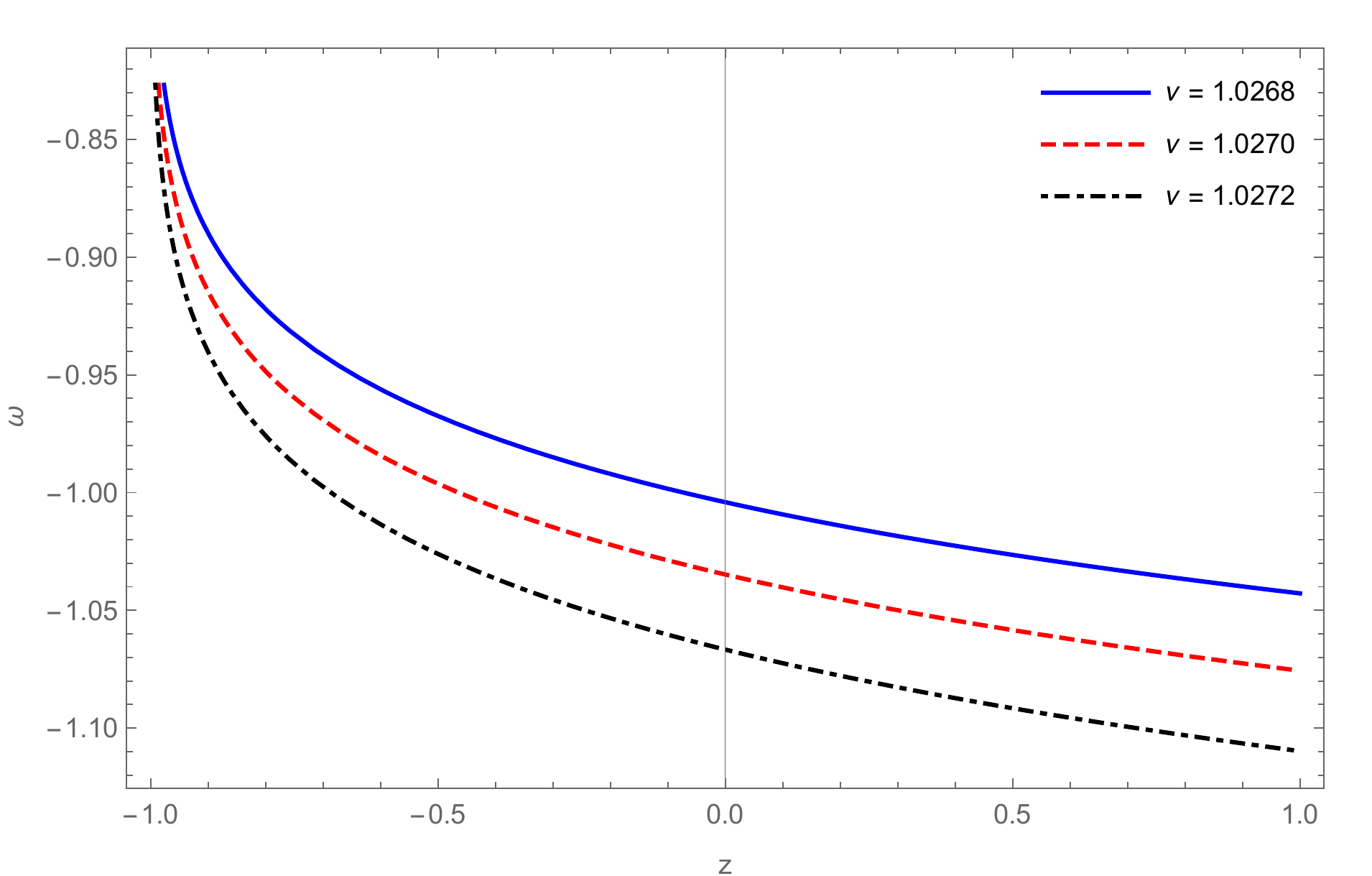}
\caption{Effective energy density (left panel) and EoS parameter (right panel) in redshift. The parameter scheme: $ \alpha=0.21, \lambda=315, \gamma=2$.}
\label{FIG3}
\end{figure}

The behaviour of these dynamical parameters depends on the parameters $\alpha$, $\nu$, $\lambda$ and $\gamma$. We plot the evolution of the energy density and EoS parameter versus the redshift by using the Eqs. \eqref{17} and \eqref{19} respectively for the dark energy sector. As shown in Fig. \ref{FIG3} (right panel), it has been realized that the EoS parameter ($\omega$) for the present epoch is less than $-1$ for the considered value of $\alpha$, $\lambda$, $\gamma$ and three different choices of $ \nu=1.0268, 1.0270, 1.0272$. Fig.\ref{3} (right panel) shows that for $ \nu=1.0268$, the EoS parameter behaves as a concordance $\Lambda$CDM phase whereas for $ \nu=1.0270, 1.0272$, it shows the phantom-like phase at present epoch. At late time, all the curves are approaching to quintessence region. As shown in Fig. \ref{3} (left panel), the effective energy density remains positive throughout the evolution. Both the energy density and EoS parameter respectively increases gradually in a positive and negative phase. The present value of the EoS parameter for this model has been noted in Table-\ref{TABLE I}  below. We wish to mention here that there is no cosmological observational value available
at present for the energy density and the value obtained for EoS parameter are within the cosmological observations.

\begin{table}[H] 
\caption{ We illustrate the present value of Hubble parameter, $H(z_0)$; Deceleration parameter, $q(z_0)$; and EoS parameter; $\omega(z_0)$ for three different parameter values of $\nu$ for both the examples.} 
\begin{center}
  \begin{tabular}{|l|l|l|l|l|l|l|l|} 
    \hline
    \multicolumn{1}{|c|}{Parameter} &
      \multicolumn{3}{c|}{First example: ~~$f(T)= T + \beta T^{2}$} &
      \multicolumn{3}{c|}{Second example: ~~$f(T)=\left(\frac{T}{\lambda}\right)^{\gamma}$} &
      \multicolumn{1}{c|}{Observations} \\
      \hline
    &$\nu= 1.0268$ & $\nu=1.0270$ & $\nu=1.0272$ & $\nu=1.0268$  &$\nu=1.0270$  &$\nu=1.0272$  & \\
    \hline
    $H(z_{0})Kms^{-1}Mpc^{-1}$ & $74$ & $73.22$ & $72.41$ &$74$  &$73.22$  & $72.41$ & $73.5\pm 1.4$ \cite{Reid19} \\
    \hline
    $q(z_{0})$ & $-1.0758$ &$ -1.0763$ & $-1.0769$ &$-1.0758$  &$-1.0763$  &$-1.0769$  &$ -1.08 \pm 0.29 $  \cite{Camarena20}  \\
    \hline
    $\omega(z_{0})$ &$ -0.93$ &$ -0.99$ & $-1.05$ &$ -1$ & $-1.02$ & $-1.05$ &$ -1.006\pm .045 $  \cite{Aghanim20}\\
    \hline
  \end{tabular}
  \end{center}
\label{TABLE I}
\end{table}

Based on the results obtained, we can infer that the form considered for the function, $f(T)$ resulted mostly phantom behaviour at present time on its evolutionary aspects. 

\section{Stability analysis in scalar perturbation approach }\label{SEC IV}
The stability analysis of the models would be performed using the scalar perturbation approach. This has become inevitable since many assumptions are made to obtain the evolutionary behaviour of the Universe and it would become difficult to assess the degree of generality of these assumptions. So, to establish the result, there is a need to investigate the qualitative features of the field equations.  Here, we shall undertake the linear homogeneous and isotropic perturbations to investigate the stability of the cosmological solutions of the $f(T)$ theory obtained in SEC-\ref{SEC III}. We shall describe the perturbations of Hubble parameter and energy density \cite{Wu12,Izumi13,Golovnev18} with the perturbation geometry functions $\delta(t)$  and matter  functions $ \delta_{m}(t)$ as,
\begin{equation}\label{20}
H(t)\rightarrow H(t)(1+\delta)  \hspace{3cm}    \rho(t)\rightarrow \rho(t)(1+\delta_{m}),
\end{equation}

The linear perturbation of the function $f$ and its derivatives are respectively $ \delta f = f_{T}\delta T$ and $\delta f_{T}= f_{TT}\delta T$ with the perturbation of torsion scalar, $T=-6H^{2}=-6H^{2}(1+\delta(t))^{2}=T(1+2\delta(t))$. Now, using the perturbative approach in the equivalent FRW  Eq. \eqref{9} background, we obtain

\begin{equation}\label{21}
-T(1+f_{T}-12H^{2}f_{TT})\delta=\kappa^{2}\rho \delta_{m},
\end{equation}

Eq.\eqref{21} describes the relationship between matter and geometric perturbation and also the perturbed Hubble parameter. However, Eq.\eqref{21} is insufficient to obtain the analytical expression for the perturbations functions. So, the perturbation continuity equation is most appropriate to use here, which can be expressed as,
\begin{equation}\label{22}
\dot\delta_{m}(t)+3H(1+\omega)\delta(t)=0
\end{equation}
which on simplification yields, 
\begin{equation}\label{23}
\delta(t)=\frac{1}{2T}. \frac{T+2Tf_{T}-f}{1+f_{T}+2Tf_{TT}} \delta_{m}(t)
\end{equation}
and subsequently the differential equation \eqref{22} becomes,
\begin{equation}\label{24}
\dot\delta_{m}(t)+\frac{3H}{2T}(1+\omega)\frac{T+2Tf_{T}-f}{1+f_{T}+2Tf_{TT}} \delta_{m}(t)=0
\end{equation}
Using the separation of variable approach for the first order ordinary differential equation in $\delta_{m} $, we find

\begin{equation}\label{25}
\delta_{m}(t)=\exp\left[-\frac{3}{2}(1+\omega)\int \frac{H}{T} \frac{T+2Tf_{T}-f}{1+f_{T}+2Tf_{TT}} dt\right]
\end{equation}
Now, with the zeroth order $tt$-component of the Friedmann equations \eqref{8} with \eqref{9} and the continuity equation  for a perfect fluid, one can get
\begin{equation}\label{26}
\dot{H} =\frac{T+2Tf_{T}-f}{4(1+f_{T}+2Tf_{TT})}(1+\omega) 
\end{equation}
From Eqs. \eqref{25} and \eqref{26}, we can get the expression for $\delta_m$ as, 
\begin{equation}\label{27}
\delta_{m}(t)=\exp\left[\int \frac{\dot{H}}{H}dt\right]=\exp \left[\int\frac{dH}{H} \right]=C_{1}H
\end{equation}

where $C_{1}$ is an integration constant and at the present moment, $C_{1}=\frac{\delta_{m}(t_{0})}{H_{0}}$ is obtained. Subsequently, we can get $\delta(t)$ as, 
\begin{equation}\label{28}
\delta(t)=-\frac{C_{1}}{3(1+\omega)}\frac{\dot{H}}{H}
\end{equation}

\begin{figure}[H]
\centering
\includegraphics[width=85mm]{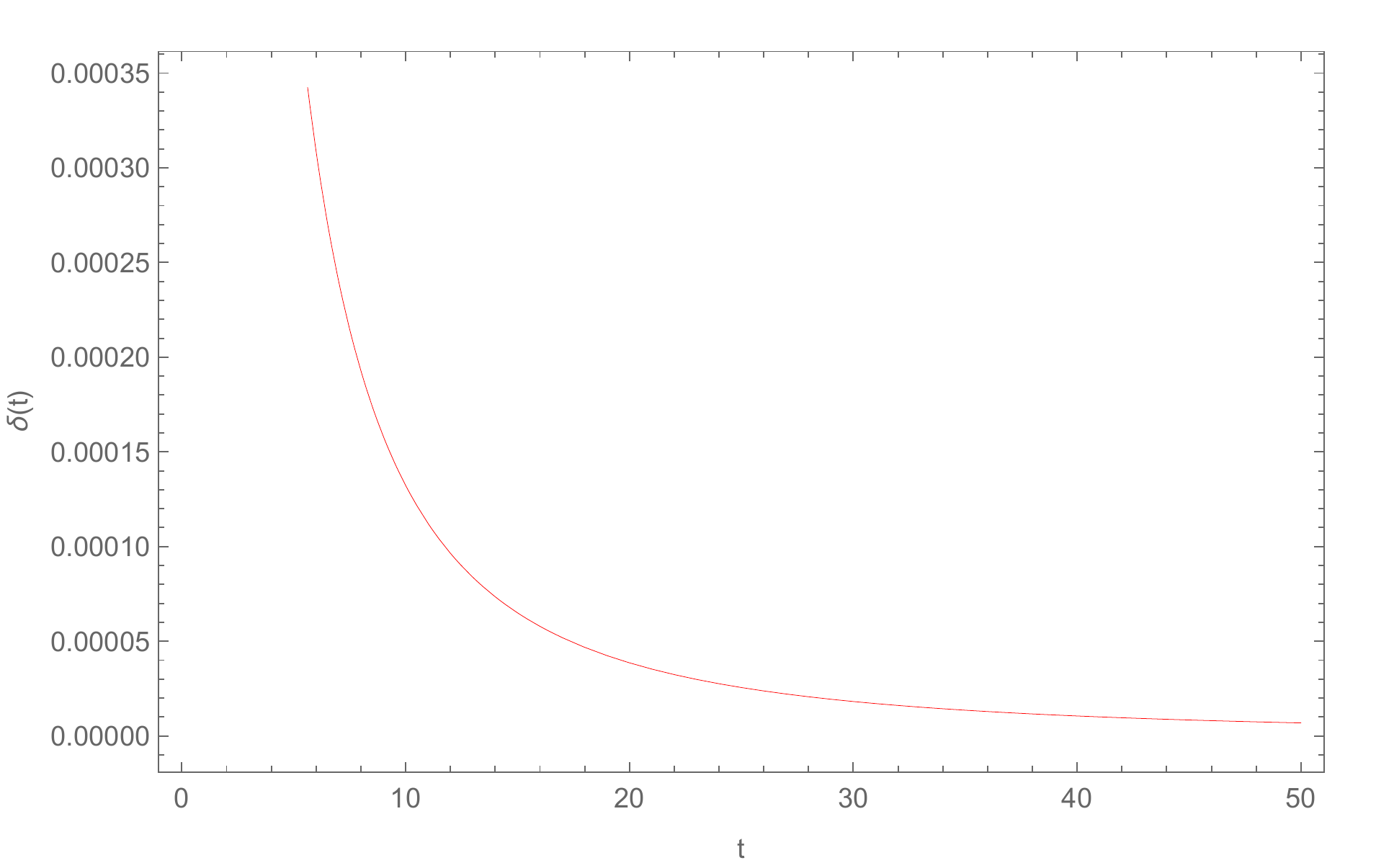}
\includegraphics[width=85mm]{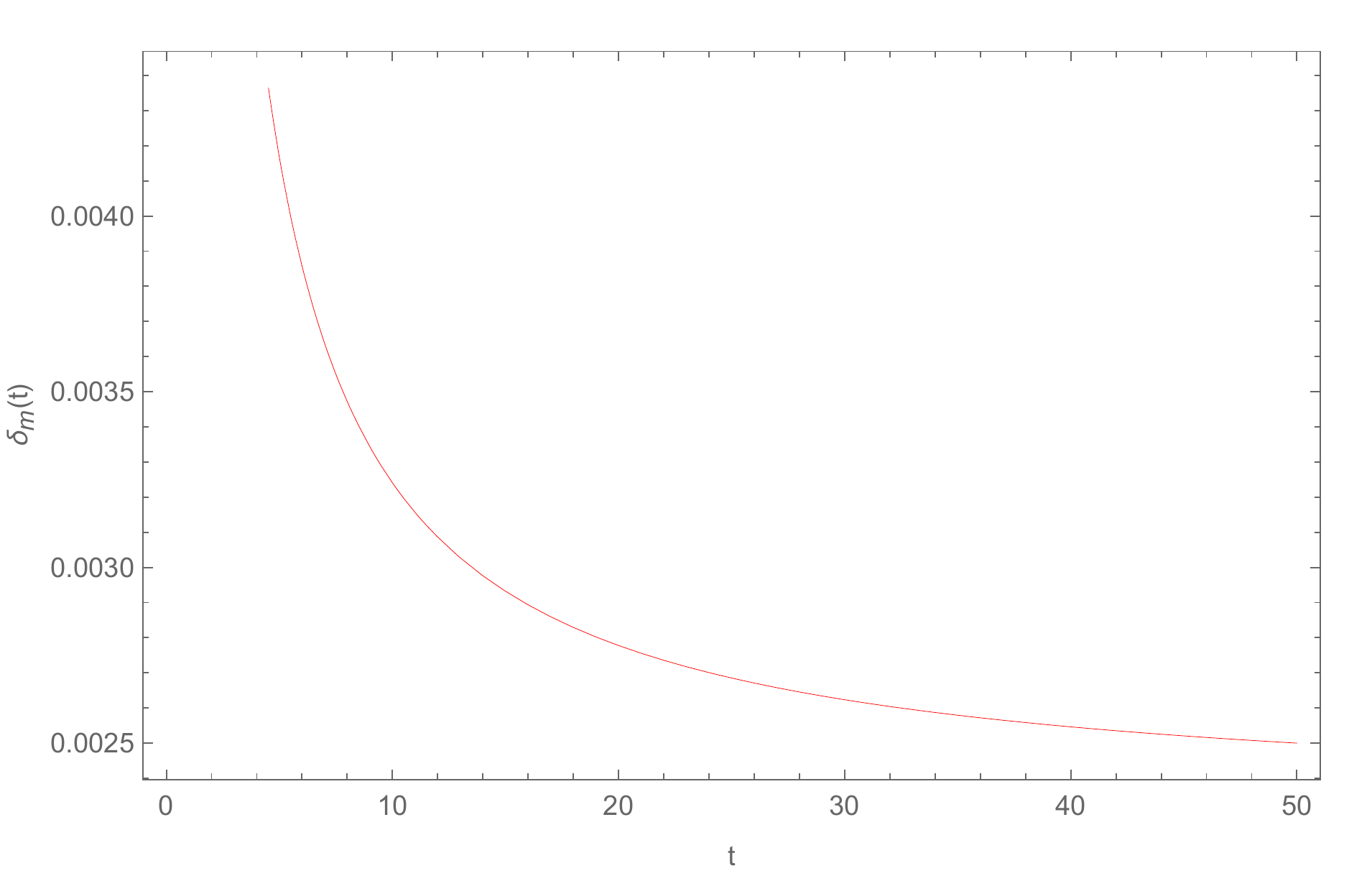}
\caption{Hubble perturbation $\delta(t)$  and matter perturbation  $\delta_{m}(t)$ parameter  in cosmic time, $\left[f(T)=T+\beta T^{2}\right]$.}
\label{FIG4}
\end{figure}

\begin{figure}[H]
\centering
\includegraphics[width=85mm]{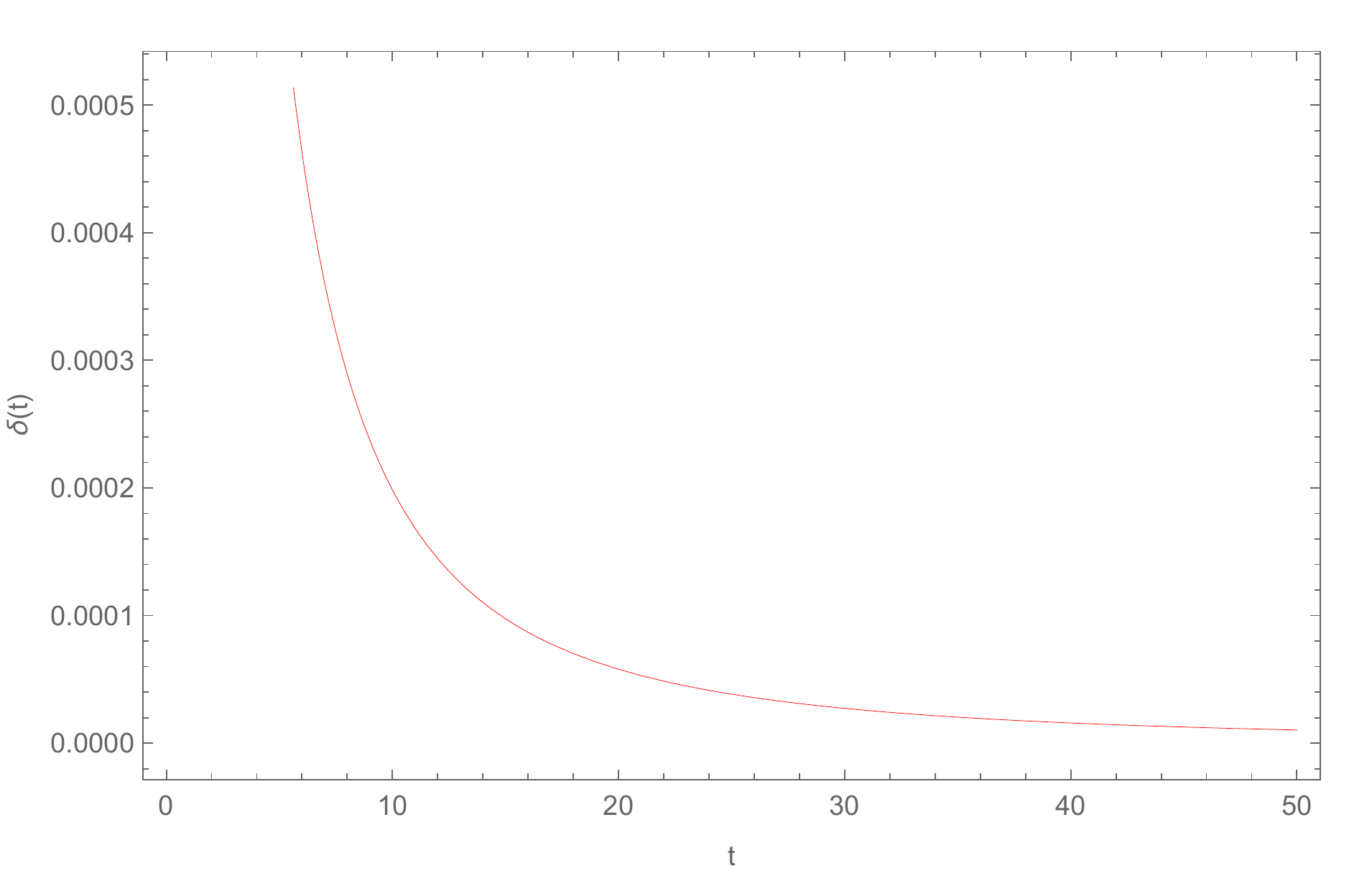}
\includegraphics[width=85mm]{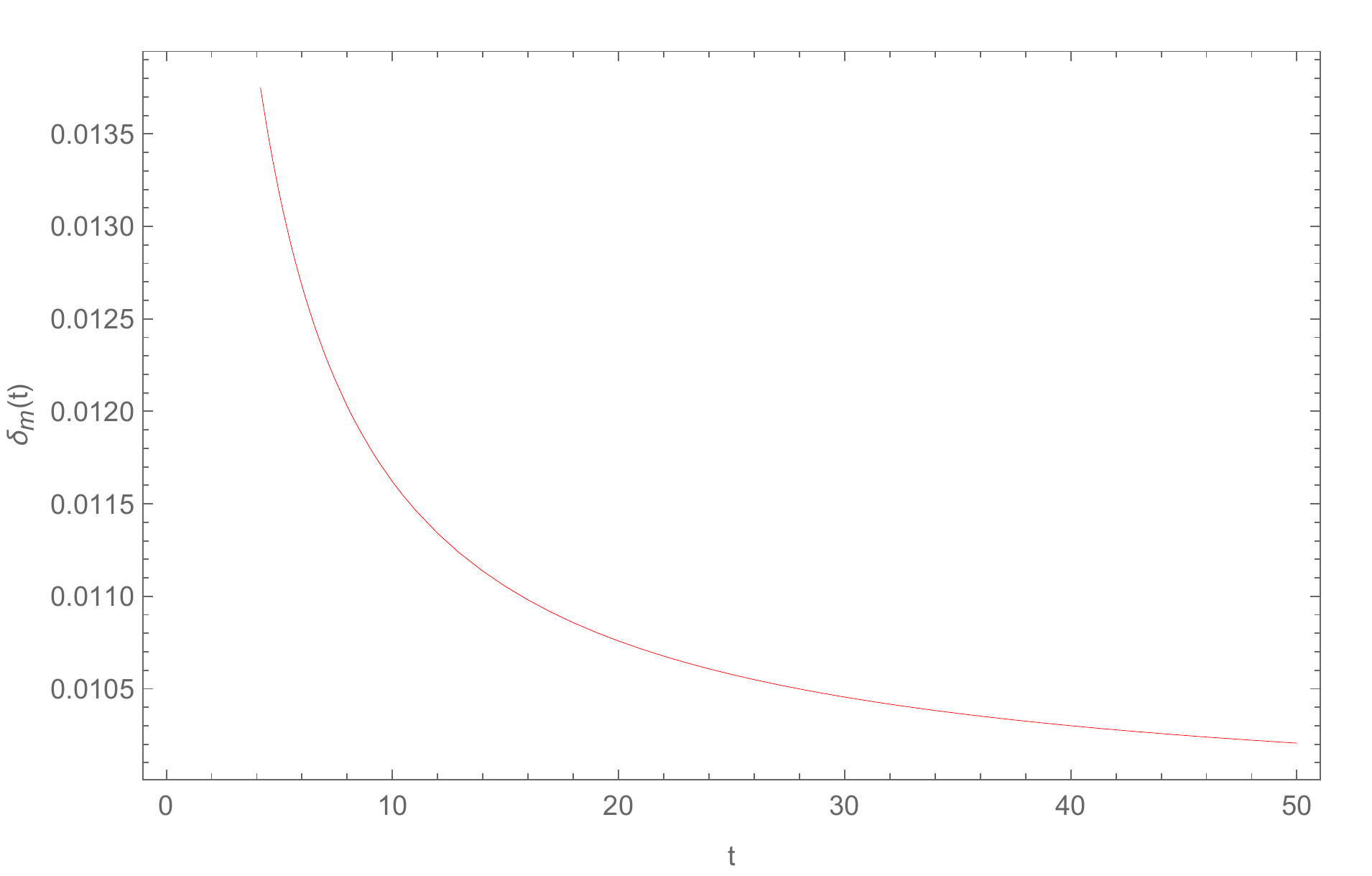}
\caption{Hubble perturbation $\delta(t)$  and matter perturbation  $\delta_{m}(t)$ parameter  in cosmic time, $\left[f(T)=\left(\frac{T}{\lambda}\right)^{\gamma}\right]$}.
\label{FIG5}
\end{figure}

From this simplification, we have seen that both $\delta_m(t)$ and $\delta(t)$ can be expressed in cosmic time. So, for both the examples discussed in SEC-\ref{SEC III}, we have plotted the behaviour of perturbed Hubble parameter and energy density. FIG. \ref{FIG4} is for the function $f(T)=T+\beta T^2$ and FIG. \ref{FIG5} for $f(T)=\left(\frac{T}{\lambda}\right)^{\gamma}$. For both the examples, $\delta(t)$ and $\delta_m(t)$ are decaying over time and at late times approaching to zero. Hence we can say both the models are stable under the scalar perturbation approach.

\section{stability analysis in dynamical system approach} \label{SEC V}

In the previous section, we have seen the stability behaviour of the models. Further to strengthen the stability behaviour of the models, in this section, we perform the dynamical system analysis.  The focus is mainly on the late time stable solutions to perform this phase-space and stability analysis. The dynamical system to be transformed into the autonomous form $X^{\prime}=f(X)$,  \cite{Bahamonde18,Mirza17}, where $X$ is the column vector constituted by suitable auxiliary variables and $f(X)$ be the corresponding column vector of the autonomous equations. The prime denotes derivative with respect to $N = ln a$. Now, the critical points can be extracted by satisfying $X^{\prime} = 0$. Then the eigenvalues for each critical points would be obtained, which would determine the stability of the cosmological models.  It has been considered that the energy components in the Universe are only dust matter  $\rho_{m}$ and radiation $\rho_{r}$. Thus
\begin{equation}
\rho = \rho_{m}+\rho_{r}, \hspace{0.9cm}       p = \frac{1}{3}\rho_{r} \label{30}
\end{equation}
 
The dimensionless phase space variables are incorporated to build an autonomous dynamical system with the help of Eqs. \eqref{7}-\eqref{8},
\begin{equation}
\Omega_{de}\equiv x= \frac{2Tf_{T}-f}{6H^{2}}  \hspace{1.5cm} \Omega_{r}\equiv y=\frac{8\pi G \rho_{r}}{3H^{2}}, \label{31}
\end{equation}
                                                      
where $\Omega_{de}$ and $\Omega_{r}$ respectively represent the  dimensionless energy density parameter of dark energy sector and radiation. Eqs. \eqref{7} and \eqref{31} provide the density parameter of the matter as, $\Omega_{m}=1-x-y $. Now,
\begin{eqnarray}
x^{\prime}&=&-\frac{(3x-3-y)(f_{T}+2Tf_{TT}+x)}{1+f_{T}+2Tf_{TT}} \label{32} \\
y^{\prime}&=&-\frac{4y(1+f_{T}+2Tf_{TT})+y(3x-3-y)}{1+f_{T}+2Tf_{TT}} \label{33}
\end{eqnarray}

Also the effective EoS parameter  and deceleration  parameter  can be expressed as in form of $x$ and $y$
\begin{eqnarray}
\omega_{eff}&=&-1-\frac{3x-3-y}{3(1+f_{T}+2Tf_{TT})} \label{34} \\
q&=& -1-\frac{3x-3-y}{2(1+f_{T}+2Tf_{TT})} \label{35}
\end{eqnarray}

Having the above general analysis in mind, one can
proceed and study the detailed dynamics of the stability of the cosmological model governed by a specific $f(T)$ model. Now, a dynamical system for the first example, $f(T)=T+\beta T^{2}$ can be obtained as 
\begin{eqnarray}
x^{\prime}&=& \frac{x(3x-3-y)}{1-2x} \label{36} \\
y^{\prime}&=& \frac{y(1-5x-y)}{1-2x} \label{37}
\end{eqnarray}

In order to extract the dynamical properties of the
above autonomous system, one solves the combined equations $x^{\prime}=0$ and $y^{\prime}=0$. The critical points obtained are $(0,0)$, $(0,1)$ and $(1,0)$, respectively labelled as $A$, $B$, and $C$. Using the Jacobian matrix of a system at each critical point, we obtained two eigenvalues. The stability nature of each critical point has been described in Table-\ref{TABLE II}. From Eqs. \eqref{34} and \eqref{35} effective  EoS  and deceleration parameters can be obtained in the form of $x$ and $y$ as,
\begin{eqnarray}
\omega_{eff}&=& -1-\frac{3x-3-y}{3(1-2x)} \label{38} \\
q&=& -1-\frac{3x-3-y}{2(1-2x)} \label{39}
\end{eqnarray}

\begin{table}[H]
\caption{Eigen values and Critical points for the dynamical system corresponding to Model-I} 
\centering 
\begin{tabular}{c c c c c c c c c} 
\hline\hline 
Critical Point & $\Omega_{m}$ & $\Omega_{r}$ & $\Omega_{de}$ & $\omega_{eff}$ & q & $\lambda_{1}$ & $\lambda_{2}$ & Stability \\ [0.5ex] 
\hline\hline 
$A(0, 0) $ & $1$ & $0$ & $0$ & $0$ & $0.5$ & $-3$ & $-1$ & Stable node\\
\hline
$B(0, 1)$ & $0$ & $1$ & $0$ & $\frac{1}{3}$ & $1$ & $-4$ & $1$ & Unstable saddle\\
\hline
$C(1, 0)$ & $0$ & $0$ & $1$ & $-1$ & $-1$ & $-3$ & $-4$ & Stable node\\[1ex] 
\hline 
\end{tabular}
\label{TABLE II}
\end{table}

Point $A$ corresponds to a matter-dominated solution, since at this point, one can easily derive that $\Omega_{m}=1$. Then one can study the stability of this solution by calculating the eigen values of the above-linearised system.
They are found to be $\lambda_{1}=-3$ and $\lambda_{2}=-1$. Hence this critical point is a stable node due to the negative eigen values, which means that the  Universe is in decelerated phase and from this critical point $A$, $\omega_{eff}=0$. Point B corresponds to a radiation-dominated solution, since it exhibits $\Omega_{r} = 1$. One can further calculate the eigen values at this point as $\lambda_{1}=-4$ and $\lambda_{2}=1$. Hence this critical point is unstable saddle due to the presence of negative and positive eigen values and this critical point shown decelerated radiation phase. Point $C$ corresponds to the solution dominated by the $f(T)$ contribution. In this solution, both $\Omega_{m}=0$ and $\Omega_{r}=0$ are vanishing. Moreover, it is easy to read that $\omega_{eff}=-1$ and both the eigen values of this critical are negative, shown in Table-\ref{TABLE II}. Therefore, the solution of point $C$ is of cosmological interest in explaining the present acceleration of our Universe.\\

\begin{figure}[H]
\centering
\includegraphics[width=85mm]{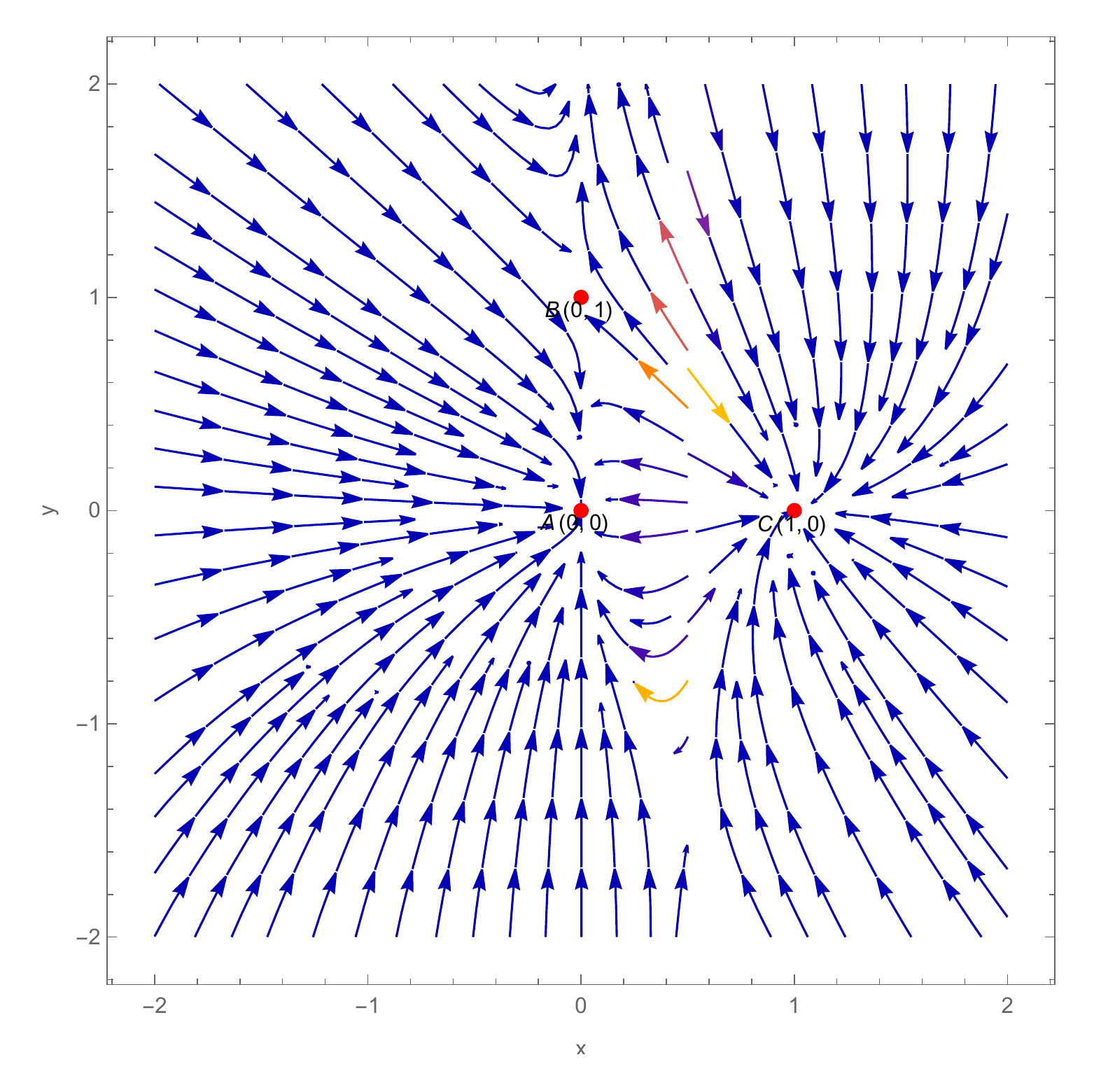}
\includegraphics[width=85mm]{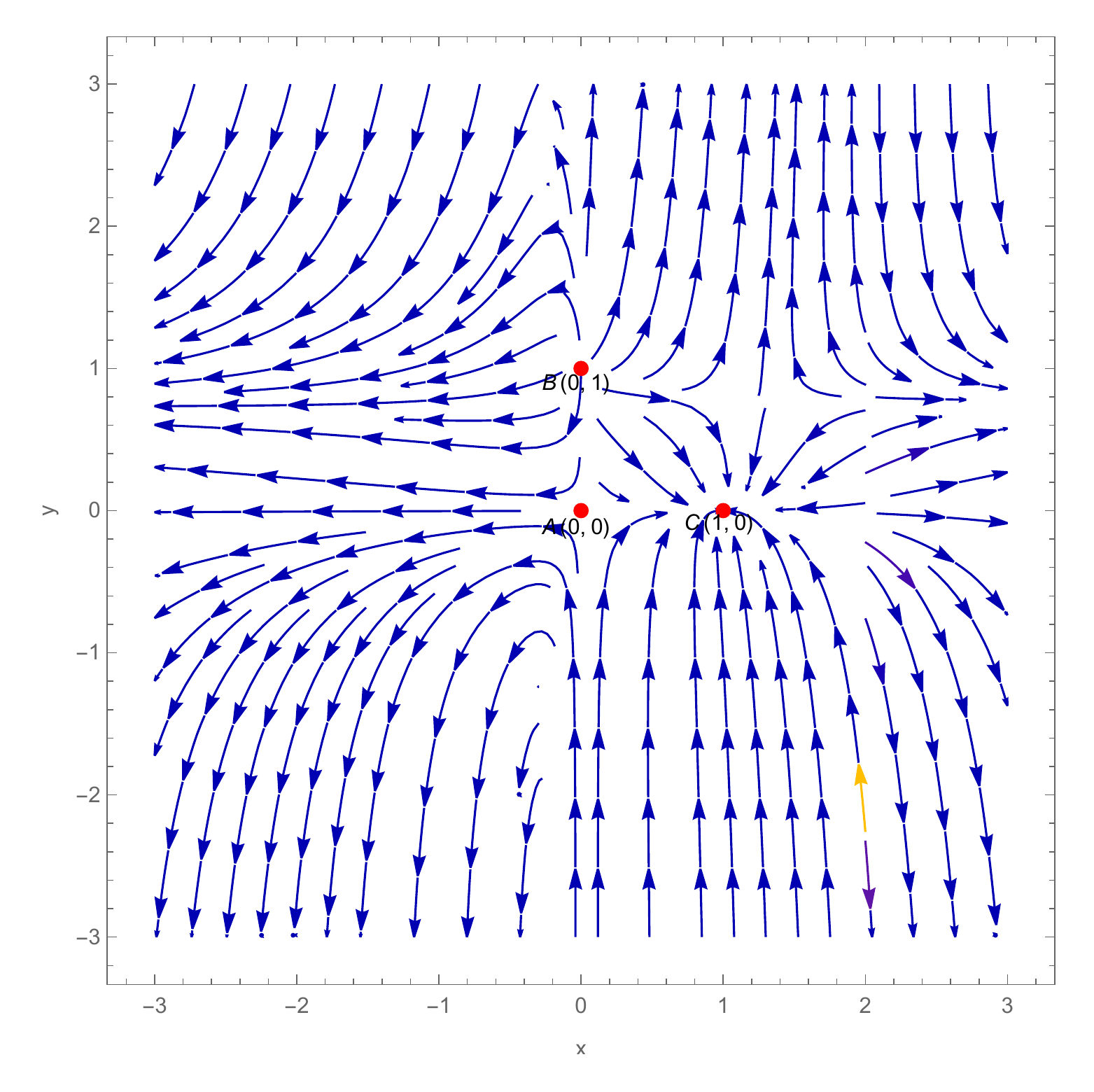}
\caption{ Phase-space trajectories on the $x-y$ plane for model-I (left panel) and model-II (right panel ) .}
\label{FIG6}
\end{figure}

In FIG. \ref{FIG6} (left panel) shows that the trajectories critical point $B(0,1)$ move away from the fixed point, so point $B$ is unstable, while they move towards  for critical points $A(0,0)$ and  $C(1,0)$, hence critical points  $A$ and $C$ behave as a stable point.

Similarly, for the second example, $f(T)=\left(\frac{T}{\lambda}\right)^{\gamma}$, the autonomous system can be obtained using Eqs. \eqref{32} and \eqref{33}, 
\begin{eqnarray}
x^{\prime}&=& -\frac{(x-x\gamma)(3x-3-y)}{1-x\gamma}\label{40}\\
y^{\prime}&=&-\frac{y(1-4xy+3x-y)}{1-x\gamma}\label{41} 
\end{eqnarray}

Also, the effective EoS and deceleration parameter respectively from Eqs. \eqref{34} and \eqref{35} can be derived in the form of $x$ and $y$ as,
\begin{eqnarray}
\omega_{eff}&=&-1-\frac{3x-3-y}{3(1-x\gamma)}\label{42} \\
q&=&-1-\frac{3x-3-y}{2(1-x\gamma)} \label{43},
\end{eqnarray}

where $\gamma$ be the model parameter and for $x^{\prime}=0$ and $y^{\prime}=0$, the critical points and its corresponding eigenvalues are obtained as in  Table-\ref{TABLE III}. 

\begin{table}[H]
\caption{Eigen values and Critical points for the dynamical system corresponding to Model-II} 
\centering 
\begin{tabular}{c c c c c c c c c} 
\hline\hline 
Critical Point & $\Omega_{m}$ & $\Omega_{r}$ & $\Omega_{de}$ & $\omega_{eff}$ & q & $\lambda_{1}$ & $\lambda_{2}$ & Stability \\ [0.5ex] 
\hline\hline 
$A(0, 0) $ & $1$ & $0$ & $0$ & $0$ & $0.5$ & $\frac{3}{2}$ & $-1$ & Unstable Saddle\\
\hline
$B(0, 1)$ & $0$ & $1$ & $0$ & $\frac{1}{3}$ & $1$ & $1$ & $ \ \ 2$ & Unstable node\\
\hline
$C(1, 0)$ & $0$ & $0$ & $1$ & $-1$ & $-1$ & $-8$ & $-\frac{3}{4}$ & Stable node\\[1ex] 
\hline 
\end{tabular}
\label{TABLE III}
\end{table}

Table-\ref{TABLE III} shows the nature of critical points for model parameter $\gamma=0.5$. Using the above value of $\gamma$, the behaviour of critical  points $A$, $B$ and $C$ are unstable saddle, unstable node and stable node respectively.
\begin{itemize}
\item Point $A$ denotes matter-dominated phase $\Omega_{m}=1$ and the corresponding eigenvalues are positive and negative, that means behaviour of this critical point is an unstable saddle and the Universe has shown decelerated behaviour.
\item Point $B$ denotes the radiation dominated phase    $\Omega_{r}=1$. Both the eigenvalues of this point are positive, which means the behaviour of this critical point is an unstable node and the Universe shown decelerated behaviour for this radiation-dominated phase.
\item Point $C$, $\omega_{eff}=-1$ shows the accelerated dark energy dominated Universe. In this critical point $\Omega_{r}=0$ and $\Omega_{m}=0$ and both the eigenvalues are negative, which means that the final state of our Universe is a stable phase. From Eq. \eqref{8} at this point $\rho=0$ and $p=0$, one can find easily that $\dot{H}=0$. At point $C$ corresponds to a de Sitter phase if $H\neq0$. Thus, for a given $f(T)$ model, the Universe finally enters a de Sitter phase.
\end{itemize}

The trajectories for critical points $A(0,0)$ and $B(0,1)$ move away from the fixed points, so points $A$ and $B$ are unstable, while they move towards for critical point $C(1,0)$, hence critical point behaves as a stable point as shown in 
FIG. \ref{FIG6} (right panel).

\section{Conclusion} \label{SEC VI}

Two cosmological acceleration models are derived based on the functional forms of $f(T)$ in an isotropic and homogeneous background. Both the models are showing phantom behaviour at the present epoch irrespective of their past and future evolutionary behaviour. In late phase, it remains in the quintessence region. With suitable choices of the model parameters, the present value obtained for the Hubble parameter, deceleration parameter and the dark energy EoS parameter as given in Table-\ref{TABLE I}. All the values of the cosmological parameters obtained are within the range of cosmological observations results. The acceleration of the model can be achieved from the behaviour of the deceleration parameter that remains negative entirely in the range, $z\approx[-1,1]$.\\ 

In the scalar perturbation approach of the stability analysis, the functions $\delta_{m}(t)\propto H $ and $\delta(t)\propto \frac{\dot{H}}{H}$. Both the functions are Hubble parameter dependent. Since the present value of the Hubble parameter is within the range of the observational results, we can expect the stable behaviour of the functions $\delta$ and $\delta_m$. Again, the Hubble parameter depends on the function $f(T)$. Therefore, the simplified form of the perturbation functions could be obtained. For both the examples, the illustrative graphs of the parameters [FIG. \ref{FIG4}- FIG. \ref{FIG5}] show the convergence of both $\delta(t)$ and $\delta_{m}(t)$ at the late times indicating the stable models. One can note that whenever $ t\rightarrow +\infty, ~~\delta(t)\rightarrow 0$ and $\delta_{m}(t)\rightarrow K$, $0<K<1$.\\

The dynamical system analysis has been used to investigate the stability of the models. In model-I critical points $A(0,0)$ and $B(0,1)$ are showing the decelerated phase of the Universe and point $C(1,0)$ represents the accelerated behavior of the Universe and at this point $C$ the effective EoS parameter is $-1$. The details of critical points are mentioned in Table-\ref{TABLE II}. In the second example, we take $\gamma=0.5$. This model is showing the unstable behavior at critical points $A(0,0)$, $B(0,1)$  and stable at the point $C(1,0)$. Points $A$ and $B$ are showing the decelerating behaviour of the Universe whereas point $C$ is showing the accelerating behavior. The details of critical points for second example are mentioned in Table-\ref{TABLE III}. Phase space trajectories for model-I and mode-II are plotted as in FIG. \ref{FIG6} left panel and right panel, respectively. This FIG. \ref{FIG6} shows both stable and unstable behavior of critical points. The stability details are mentioned in Table-\ref{TABLE II} and Table-\ref{TABLE III} for model-I and model-II, respectively.\\

Finally, we can conclude that accelerating stable cosmological models can be achieved by including torsion in place of curvature in the action formula. Also it depends on the form of $f(T)$ chosen in the model. These results further strengthen the viability of teleparallel gravity in addressing the issue of late-time cosmic acceleration.

\section*{Acknowledgement}
LKD and SVL acknowledge the financial support provided by University Grants Commission (UGC) through Junior Research Fellowship (UGCRef. No.: 191620180688 (LKD), 191620116597(SVL)) to carry out the research work. BM and SKT acknowledge the support of IUCAA, Pune (India) through the visiting associateship program.

\end{document}